\documentclass[prl,twocolumn,showpacs,preprintnumbers,amsmath,amssymb,superscriptaddress,floatfix,showkeys]{revtex4}

\usepackage{graphicx}
\usepackage{dcolumn}
\usepackage{bm}
\usepackage{color} 
\usepackage{amsfonts}
\usepackage{amsmath}
\usepackage{amsthm}
\usepackage[version=3]{mhchem}

\begin{document} 
 
\title{Controlling the microstructure and phase behavior of confined soft colloids by active interaction switching}

\author{Arturo Moncho-Jord\'a} 
\affiliation{Departamento de F\'{\i}sica Aplicada and Instituto Carlos I de F\'{\i}sica Te\'orica y Computacional, Universidad de Granada, Campus Fuentenueva S/N, 18071 Granada, Spain.}
\email{moncho@ugr.es}
%\affiliation{Instituto Carlos I de F\'{\i}sica Te\'orica y Computacional, Facultad de Ciencias, Universidad de Granada, Campus Fuentenueva S/N, 18071 Granada, Spain.}
\author{Joachim Dzubiella} 
\affiliation{Applied Theoretical Physics-Computational Physics, Physikalisches Institut, Albert-Ludwigs-Universit\"at Freiburg, Hermann-Herder Stra{\ss}e 3, D-79104 Freiburg, Germany.}
\affiliation{Cluster of Excellence livMatS @ FIT - Freiburg Center for Interactive Materials and Bioinspired Technologies, Albert-Ludwigs-Universit\"at Freiburg, Georges-K{\"o}hler-Allee 105, D-79110 Freiburg, Germany}
%\affiliation{Research Group for Simulations of Energy Materials, Helmholtz-Zentrum Berlin, D-14109 Berlin, Germany}
\email{joachim.dzubiella@physik.uni-freiburg.de}

\date{\today}
 
\begin{abstract}
We explore the microstructure and phase behavior of confined soft colloids which can {\it actively} switch their interactions at a predefined kinetic rate. For this, we employ a reaction-diffusion approach based on a reactive dynamical density-functional theory (R-DDFT) and study the effect of a binary (two-state) switching of the size of colloids interacting with a Gaussian pair potential. The rate of switching interpolates between a near-equilibrium binary Gaussian mixture at low rates and an effective monodisperse liquid for large rates, hence strongly affecting the one-body density profiles, adsorption, and pressure at confining walls. Importantly, we demonstrate that sufficiently fast switching impedes the phase separation of an (in equilibrium) unstable liquid, allowing the dynamic control of the degree of mixing/condensation and local micro-structuring in a cellular confinement by tuning the switching rate. \end{abstract} 

\maketitle 
\vspace*{-0.5cm}

Active matter systems have drawn the attention of the soft matter scientific community in the recent years due to their very rich dynamic behavior. They are composed by individual particles, each of which consumes energy in order to move, react, or to produce mechanical forces. Prominent examples of active soft matter systems are self-propelled nanoparticles, contracting biopolymers such as proteins or actin filaments inside the cytoskeleton, biological cells and bacteria~\cite{Ramaswamy2010,active_materials,IDP_switching,Marchetti2013,helical_switching}, or synthetic active hydrogels~\cite{oscillating, Heuser, DNA_hydrogel,breathing}. Fascinating new dynamics have been revealed, for example, self-propelling activity has been shown to lead to motility-induced swarming, jamming, or phase-separation~\cite{Cates2015, RevModPhys88}. 

Biological activity, in particular through fuel-driven changes of molecular properties and conformations, has been made responsible for liquid-liquid phase separation and condensation in cells, with large implications for physiology and disease~\cite{Shin,Berry2018}. Living cells contain distinct sub-compartments to facilitate spatiotemporal regulation of biochemical reactions where transient microstructuring is key for function. Those biology-inspired non-equilibrium, transient morphologies bear potential for the design of novel adaptive materials~\cite{Andreas}, e.g., by harvesting switchable self-assembly and structuring~\cite{Heuser}. The microscopic origins and features of non-equilibrium structuring, however, are not well understood. Theoretical frameworks for interacting reaction-diffusion systems have been linked to microstructuring dynamics to describe non-active systems driven by chemical reactions~\cite{Glotzer1995,Lutsko2016,AndyPRE2018, Liu2020} or virus infections~\cite{corona}. However, activity controlled structuring and demixing/condensation in confinement by switching microscopic conformations and interactions has not been addressed so far. 

In this Letter, in contrast to the well-studied motile activity~\cite{Cates2015, RevModPhys88}, we investigate the effects of active conformational switching of particles on the liquid microstructuring in confinement, such as a wall or a cellular compartment. For this, we employ a reactive dynamical density functional theory (R-DDFT) put forward in other flavors previously~\cite{Glotzer1995, Lutsko2016,AndyPRE2018, Liu2020} and apply it to the binary switching of the size of soft, interacting colloids. The latter serve as a generic model for polymers and soft colloidal hydrogels~\cite{SoftColloid,Scotti} and cells~\cite{Winkler}, where size is used as the simplest interaction variable to describe a conformational change. We show that the rate of switching then interpolates between an equilibrium binary mixture at low rates to an effective monodisperse system for large rates. As a consequence, the variation of the rate substantially modifies the microstructure and phase separation of the active liquid in confinements. Hence, we demonstrate that active interaction switching dynamically controls the degree of demixing/condensation and local micro-structuring in compartmentalized situations. 

As a simple model system, mimicking, e.g., soft active hydrogels switching (or 'breathing') between two states~\cite{oscillating, Heuser, DNA_hydrogel, breathing} or responsive, conformationally switching biopolymers~\cite{Shin,IDP_switching, helical_switching}, we employ repulsive soft particles that can actively switch between two states `big' (b) and `small' (s) which only differ in the size of the particles.  The interactions are defined by soft repulsive Gaussian pair potentials~\cite{Stillinger1976}, $\beta u_{ij}(r)=\epsilon_{ij}e^{-r^2/\sigma_{ij}^2}$ ($i,j=\textmd{s},\textmd{b}$), where $r$ is the interparticle distance, $\beta=1/(k_BT)$ the inverse thermal energy, $\epsilon_{ij}>0$ denote the strength of the interactions, and $\sigma_{ij}$ represent their range (we will denote $\sigma_\textmd{bb}$ and $\sigma_\textmd{ss}$ by $\sigma_\textmd{b}$ and $\sigma_\textmd{s}$, respectively). The main feature of the Gaussian pair potentials, appropriate to describe polymers and soft colloids~\cite{SoftColloid,Bolhuis}, is that they remain finite for any interparticle distance, so two particles can interpenetrate each other. A big benefit is also that one- and two-component Gaussian systems are well understood in and out-of equilibrium and the mean-field free energy functional in DFT is quasi-exact for these systems~\cite{Louis2000,Lang_2000,Archer2001,Dzubiella2003,Archer2005,Archer2005a}. 
\begin{table}
	\caption{Parameters describing particle interactions and concentrations for three Gaussian mixtures. Systems S1 and S2 are stable whereas system S3 is unstable.}
	\label{tbl:systems}
	\begin{tabular}{ |c|c|c|c|c|c|c|c|c|c| }
		\hline
		System & $\epsilon_\textmd{bb}$ & $\epsilon_\textmd{ss}$ & $\epsilon_\textmd{bs}$ & $\sigma_\textmd{b}/\sigma_\textmd{s}$ &  $\sigma_\textmd{bs}/\sigma_\textmd{s}$ & $\rho_\textmd{T}\sigma_\textmd{s}^3$ & $\phi_\textmd{T}$ & $x$ & $k_\textmd{bs}$/$k_\textmd{sb}$ \\ 
		\hline
		S1 & 2 & 2 & 2 & 2 & 1.5 & 0.239 & 0.3 & 0.8 & 4 \\ 
		S2 & 2 & 2 & 2 & 2 & 1.5 & 0.191 & 0.45 & 0.5 & 1 \\ 
		S3 & 2 & 2 & 1.888 & 1.504 & 1.277 & 2.4 & 2.765 & 0.5 & 1\\ 
		%		S4 & 2 & 2 & 1.888 & 1.504 & 1.277 & 0.855 & 1.005 & 2.4 & 0.8 \\ 
		\hline
	\end{tabular}
\end{table}

We emphasize that this system is actually a one-component system, but since every individual particle has two states (b and s), every microstate and also average (steady-state) distributions have to be described as for a binary mixture.  Hence, every state is formed by $N_\textmd{b}$ repulsive Gaussian particles of type $\textmd{b}$ and $N_\textmd{s}$ particles of type $\textmd{s}$ contained within a volume $V$, at fixed temperature $T$.  The bulk number densities of both species are given by $\rho_i=N_i/V$ ($i=\textmd{b}, \textmd{s}$). We denote the total number density by $\rho_\textmd{T}=\rho_\textmd{b}+\rho_\textmd{s}=const$, and define a concentration ratio by $x=\rho_\textmd{s}/\rho_\textmd{T}$. Analogously, the total volume fraction of particles is given by $\phi_\textmd{T}=\phi_\textmd{b}+\phi_\textmd{s}=(\pi/6) (\rho_\textmd{b}\sigma_\textmd{b}^3+\rho_\textmd{s}\sigma_\textmd{s}^3)$. Non-equilibrium switching activity implies that particles of type $\textmd{b}$ can be converted into $\textmd{s}$ at a rate $k_\textmd{bs}$ (units of time$^{-1}$). Conversely, particles of type $\textmd{s}$ convert into $\textmd{b}$ at a rate $k_\textmd{sb}$. This `chemical' reaction process is ruled by the well-known set of first-order differential equations~\cite{Dill}
\begin{equation}
\label{kinetic_equations}
d\rho_\textmd{b}/dt=k_\textmd{sb}\rho_\textmd{s}-k_\textmd{bs}\rho_\textmd{b}, \ \ \ \ d\rho_\textmd{s}/dt=k_\textmd{bs}\rho_\textmd{b}-k_\textmd{sb}\rho_\textmd{s}
\end{equation}
The integration of these equations leads to exponential decaying time-dependent concentrations (see Supplemental Material). In particular, the equilibrium composition achieved in the limit $t \rightarrow \infty$ satisfies $\rho_{\textmd{s},\infty}/\rho_{\textmd{b},\infty}=x_{\infty}/(1-x_{\infty})=k_\textmd{bs}/k_\textmd{sb}$, where $\rho_{i,\infty}$ are the equilibrium densities. 

How is the average (steady-state) liquid microstructure of such a reactive system under a confining external potential $u_i^\textmd{ext}(\bold{r})$ ($i=\textmd{b}, \textmd{s}$) affected by the switching rate? In order to answer this question we make use of DDFT, which represents an adaptation of the equilibrium density functional theory for interacting fluids of Brownian (overdamped) particles to a non-equilibrium situation~\cite{Marconi1999,Marconi2000, Archer2004}, including the action of motile activity~\cite{Hartmut2008,Witti2011,Stark2011}. DDFT has proven its effectiveness in describing the out-of-equilibrium dynamics under constant or time dependent external potentials in many different soft matter systems and applications~\cite{Rauscher2013,Angioletti-Uberti2014,Angioletti-Uberti2018,Moncho-Jorda2019}, including the Gaussian interaction~\cite{Dzubiella2003,Archer2005,Archer2005a}. Here, we extend the DDFT formulation to active mixtures in which each component switches into the other at some fixed rate. In this process, the time evolution of the particle concentrations $\rho_i(\bold{r},t)$ are not only due to the diffusive fluxes, but also to the production and disappearance of each component. These processes occur locally, so the conversion rate of big colloids into small ones and vice versa will depend on the local concentrations of both species. Analogous to related DDFT-based reaction-diffusion approaches~\cite{Lutsko2016,AndyPRE2018,Liu2020,corona}, the governing reaction-diffusion equations read
\begin{equation}
\begin{cases}
\frac{\partial \rho_\textmd{b}(\bold{r},t)}{\partial t}=-\nabla \cdot \bold{J}_\textmd{b} + k_\textmd{sb}\rho_\textmd{s}(\bold{r},t) - k_\textmd{bs}\rho_\textmd{b}(\bold{r},t) \\
\frac{\partial \rho_\textmd{s}(\bold{r},t)}{\partial t}=-\nabla \cdot \bold{J}_\textmd{s} + k_\textmd{bs}\rho_\textmd{b}(\bold{r},t) - k_\textmd{sb}\rho_\textmd{s}(\bold{r},t)
\end{cases},
\label{RDDFT}
\end{equation}
where $\bold{J}_i=-D_i\left[\nabla \rho_i+\rho_i\nabla \beta( u_i^\textmd{ext}+\mu_i^\textmd{ex})\right]$ ($i=\textmd{b}, \textmd{s}$) are the corresponding diffusive fluxes, $D_i =k_BT/(3 \pi \eta \sigma_i )$ are the diffusion constants of both species, and $\mu_i^\textmd{ex}(\bold{r},t)=\delta F_\textmd{ex}[\{ \rho_i(\bold{r},t) \}]/\delta \rho_i(\bold{r},t)$ are the functional derivatives of the equilibrium excess free energy functional with the equilibrium density profiles replaced by the non-equilibrium ones, $\rho_i(\bold{r},t)$. Since Gaussian particles behave as a weakly correlated mean-field fluid over a surprisingly wide density and temperature range~\cite{Louis2000}, we use $F_\textmd{ex}=\frac{1}{2}\sum_{i,j=\textmd{b},\textmd{s}}\iint \rho_i(\bold{r},t) \rho_j(\bold{r^{\prime}},t) u_{ij}(|\bold{r}-\bold{r}^{\prime}|)d\bold{r}d\bold{r}^{\prime}$.

This reactive DDFT (R-DDFT) framework, Eq.~(\ref{RDDFT}), together with the explicit mean-field expression of $F_\textmd{ex}$, provides an approximate tool to calculate the non-equilibrium kinetics and micro-structure of colloids with active switching of interactions in the presence of external potentials. Details about the boundary conditions and numerical integration used to solve these equations are shown in the Supplemental Material. 
%\begin{figure}[ht!]
%	\centering
%	\includegraphics[width=0.9\linewidth]{1wall_densityprofiles_a_PRL.pdf}
%	\caption{Final steady-state density profile of big (top panel) and small (bottom panel) particles near a hard wall for different values of the exchange activity, from $a=0$ (non-active) to $a=10^4$ (System S1).}
%	\label{fig:1wall_densityprofiles_a}
%\end{figure}
\begin{figure}[ht!]
	\centering
	\includegraphics[width=1.0\linewidth]{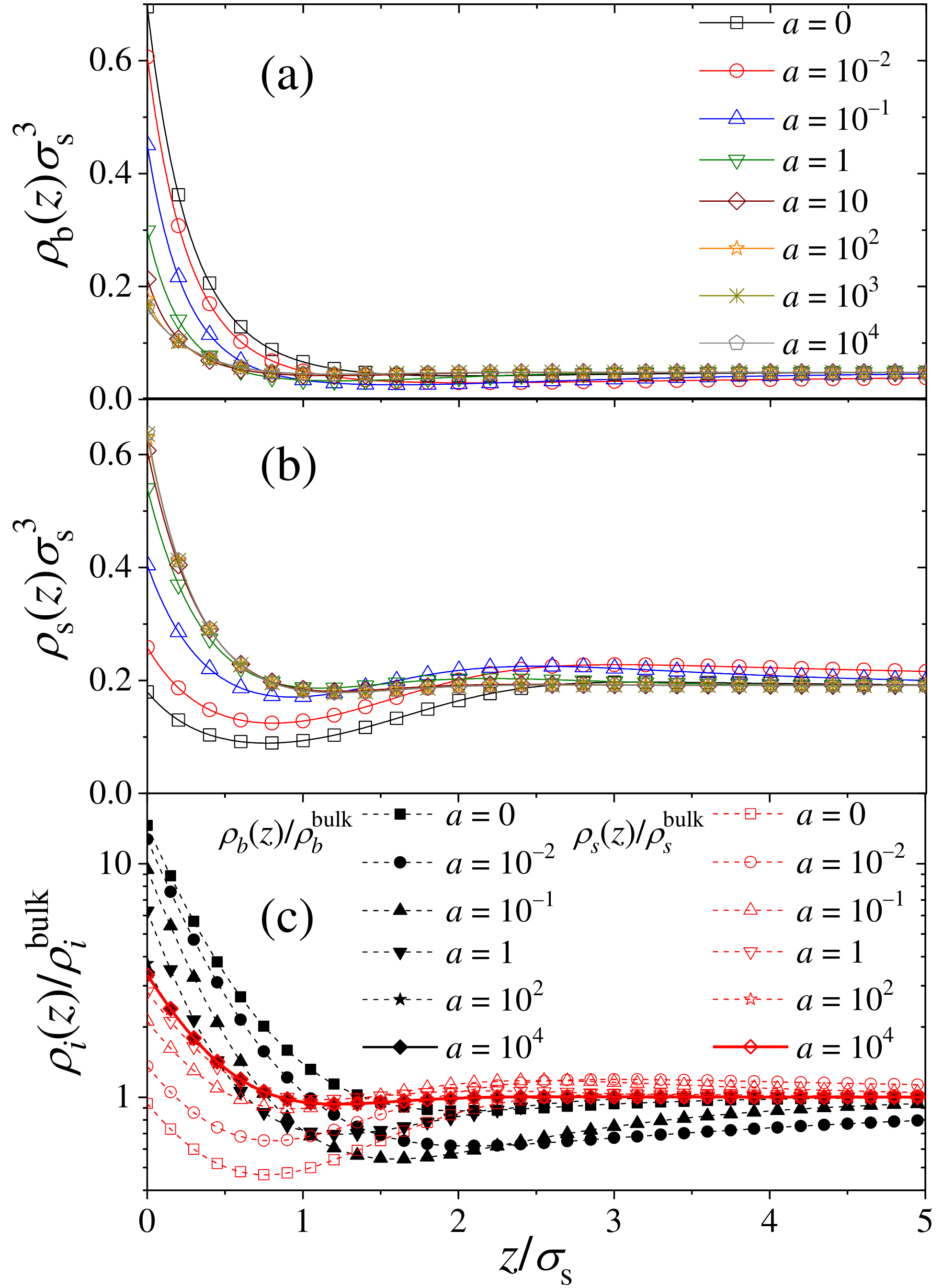}
	\caption{(a) and (b) Final steady-state density profile of big (top panel) and small (bottom panel) particles near a hard wall for different values of the exchange activity, from $a=0$ (non-active) to $a=10^4$ (system S1). (c) Normalized steady-state density profiles ($\rho_i(z)/\rho_i^\textmd{bulk}$) of big and small particles converge to the same profile (thick red line) for $a \rightarrow \infty$ (system S1).}
	\label{fig:1wall_densityprofiles_a_convergence}
\end{figure}

As first applications we investigate emblematic inhomogeneous situations in confining potentials, namely the liquid structuring at a single hard wall and the phase separation in a cellular confinement, in which the kinetic rate constants $k_\textmd{bs}$ and $k_\textmd{sb}$ for active switching fulfill the condition $k_\textmd{bs}/k_\textmd{sb}=x/(1-x)$. In this way, reference bulk concentrations remain unaltered, but the inhomogeneous properties suffer the effect of the switching. We define the normalized switching rate, or \textit{switching activity} as $a=k_\textmd{bs}\sigma_\textmd{s}^2/D_s$. For $a \ll 1$, the \ce{b <=> s} conversion rate is so slow that the time evolution of the density profiles is dominated by diffusion. In this case, any exchange at some specific location is rapidly compensated by diffusive fluxes of particles that balance the effect of the activity. It is easy to see from Eq.~(\ref{RDDFT}) that the steady-state ($\partial \rho_i/\partial t =0$) then reduces to the situation of a binary mixture in equilibrium.  Conversely, for $a \gg 1$ the exchange rate is so large that the diffusion is not fast enough to compensate its effects, so the exchange activity dominates. In this limit, we will demonstrate that the system behave in the non-equilibrium steady-state effectively as a monodisperse system of same size. For $a=1$ both processes have the same relative importance.

Table~\ref{tbl:systems} specifies the interaction parameters and particle concentrations for the three particular systems (systems S1 to S3) that will be the matter of investigation. Systems S1 and S2 are inside the stable region of the phase diagram, whereas system S3 is located in the unstable region, so it undergoes a fluid-fluid demixing in equilibrium. For all cases, we study the final steady state for active systems from $a=0$ (non-active equilibrium) to large activities $a=10^4$.

We first investigate the role that switching activity plays on the structure and adsorption of the liquid at a hard wall. We show in Figs.~\ref{fig:1wall_densityprofiles_a_convergence}(a) and (b) the final steady state for different active systems from $a=0$ to $a=10^4$ (for the non-equilibrium relaxation from a non-steady state see an example in the Supplemental Material). These results clearly show that the equilibrium density profiles ($a=0$) near the wall experience a considerable change even for relatively small values of the activity, see, for instance, how the whole curve $\rho_\textmd{s}(z)$ significantly raises up for a relatively small value of the activity $a=0.01$. Increasing the value of $a$ entails a progressive desorption of big colloids from the wall and an enhanced adsorption of small ones. 

Importantly, for growing $a \gg 1$ the system does not exhibit a temporal unbound, diverging structure of the particle concentrations. In fact, in the limit of very large activities the system finally tends to a steady-state in which the density profiles of big and small particles converge to each other. This fact can be straightforwardly confirmed in Fig.~\ref{fig:1wall_densityprofiles_a_convergence}(c), where the normalized density profiles $\rho_i(z)/\rho_i^\textmd{bulk}$ of big (black curves) and small spheres (red curves) are depicted together in the same plot for increasing values of $a$. 
%On the one hand, big spheres desorb from the wall for increasing activity while $\rho_\textmd{b}(z)/\rho_\textmd{b}^\textmd{bulk}$ develops a depletion region at $z/\sigma_s \simeq 1.5$ for $a<1$ that repletes again for $a>1$. On the other hand, the adsorption of small spheres is enhanced and the depletion region fills non-monotonously with increasing activity.
 For $a   \gtrsim 10^2$ both profiles converge to a common profile (thick red line) that is somewhat in the middle. In other words, in the limit of large switching activity, the switching system behaves as an effective monodisperse system of colloids of same interaction size.
\begin{figure}[ht!]
	\centering
	\includegraphics[width=1.0\linewidth]{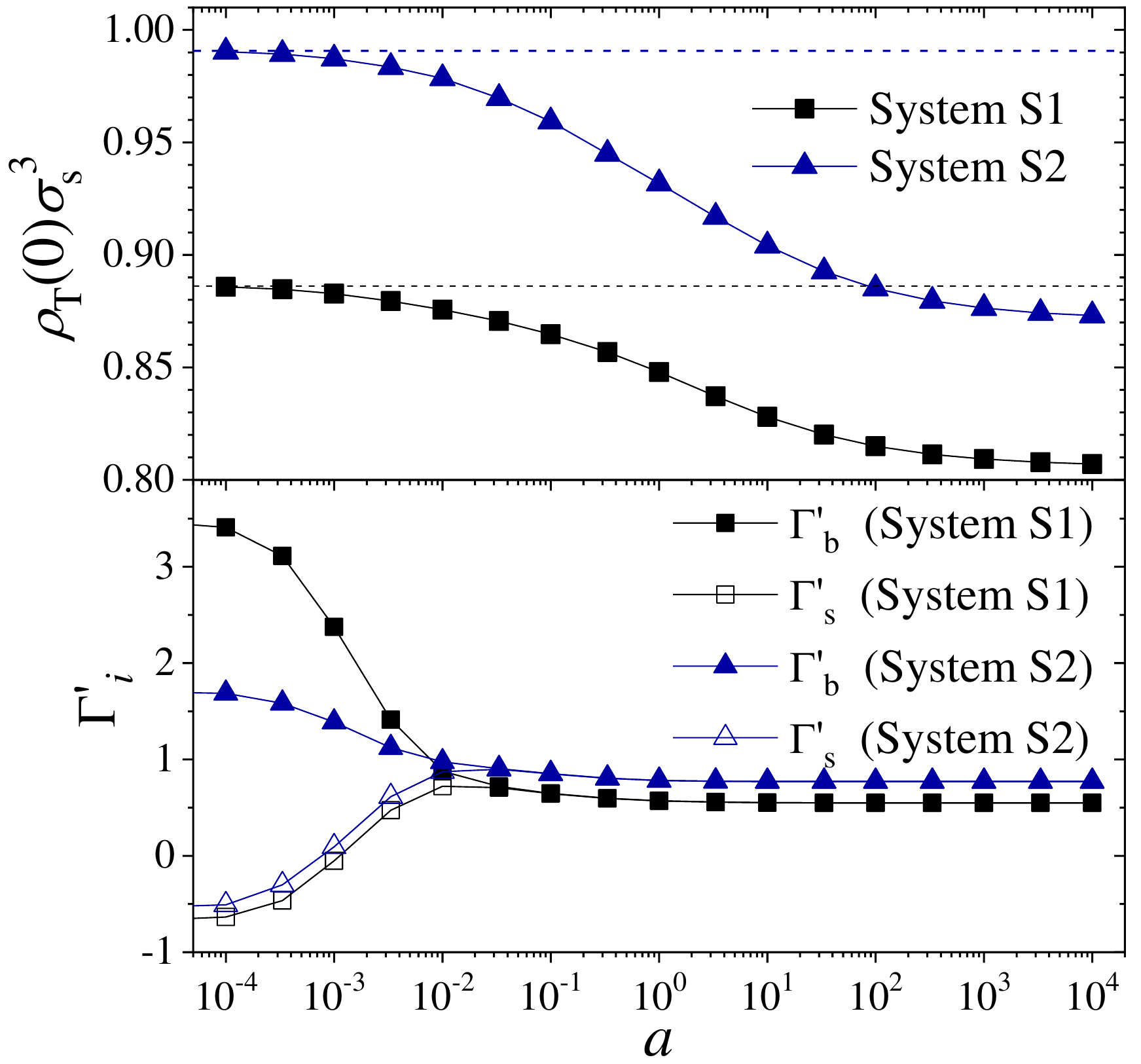}
	\caption{Top panel: Pressure exerted by a binary mixture of Gaussian particles onto a hard planar wall versus $a$ in the steady state. Dashed horizontal lines show the bulk pressure for $a=0$. Bottom panel: reduced adsorption of big and small particles onto a hard planar wall versus $a$. Black squares (system S1); blue triangles (system S2).}
	\label{fig:1wall_pressure_gamma}
\end{figure}

Activity also has important repercussions on the wall pressure of the active steady-state system. Fig.~\ref{fig:1wall_pressure_gamma} (top panel) displays $\beta P_\textmd{wall}\sigma_\textmd{s}^3=\rho_\textmd{T}(0)\sigma_\textmd{s}^3$ as a function of $a$ for systems S1 and S2. In both cases, the systems exactly satisfy the contact wall theorem for $a=0$, i.e., the pressure at the wall agrees with the bulk pressure, given by $\beta P_\textmd{bulk}=\rho_\textmd{T}+\frac{1}{2}\rho_\textmd{T}^2\pi^{3/2}\sum_{i,j=\textmd{b},\textmd{s}}x_ix_j\epsilon_{ij}\sigma_{ij}^3$ (shown as horizontal dashed lines). However, increasing the switching activity yields a systematic departure from this behavior, as the pressure at the wall decreases monotonously until it reaches the asymptotic limit for $a\rightarrow \infty$. Consequently, active interaction switching tunes the bulk pressure. 

%\begin{figure*}[htb]
%	\centering
%	\includegraphics[width=0.9\linewidth]{phase_separation1_PRL_new.pdf}
%	\caption{Steady-state density profiles for system S3 confined inside a spherical cavity (cell) of radius $R=5\sigma_\textmd{s}$, for activities from $a=0$ to $10^3$. For $a=0$ the system is in equilibrium and demixed. Increasing $a$ suppress the phase separation and induces mixing. Inset shows $\delta$ as a function of $a$.}
%	\label{fig:phase_separation1}
%\end{figure*}
\begin{figure}[ht!]
	\centering
	\includegraphics[width=1.0\linewidth]{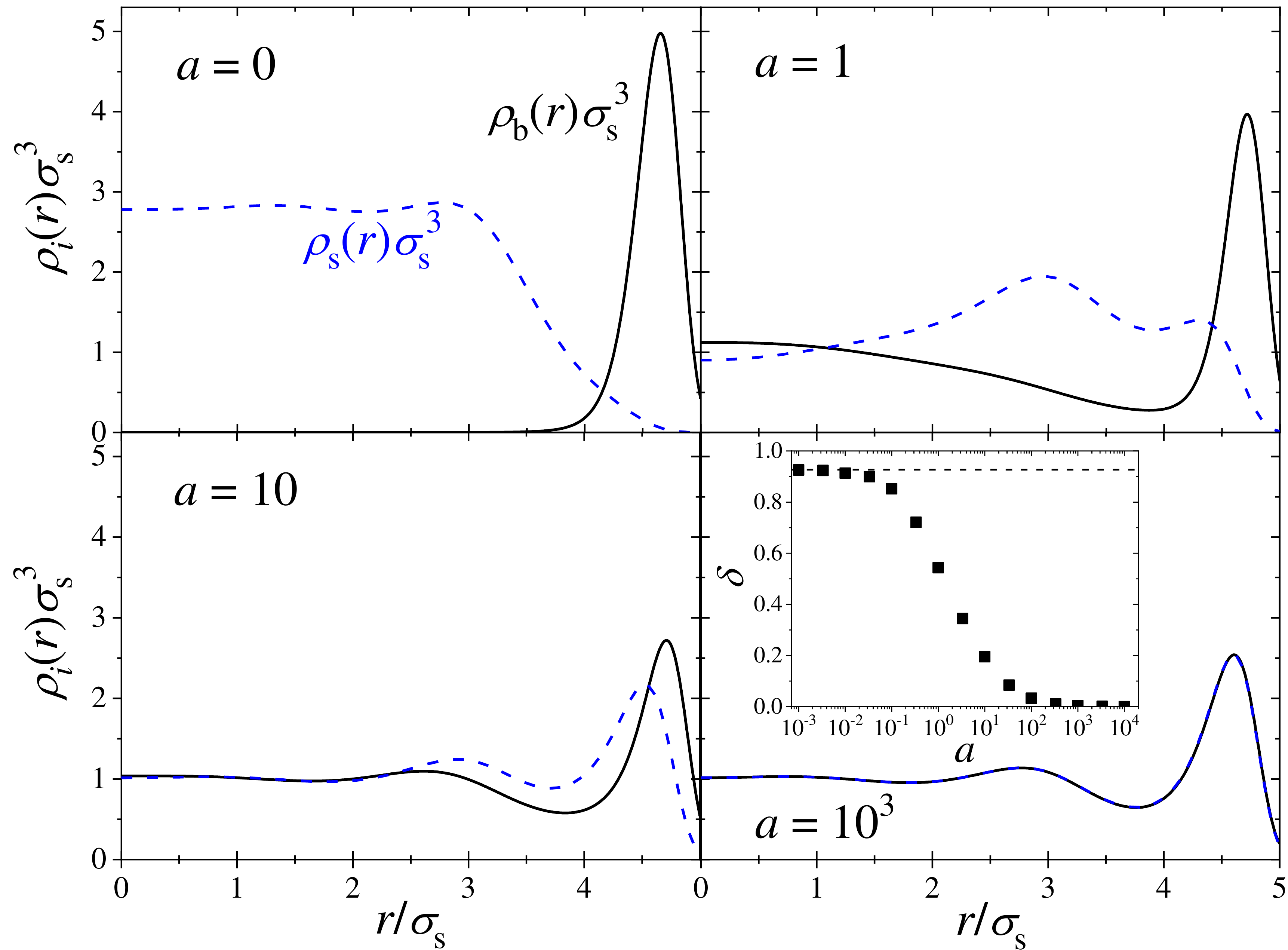}
	\caption{Steady-state density profiles for system S3 confined inside a spherical cavity (cell) of radius $R=5\sigma_\textmd{s}$, for activities from $a=0$ to $10^3$. For $a=0$ the system is in equilibrium and demixed. Increasing $a$ suppress the phase separation and induces mixing. Inset shows $\delta$ as a function of $a$.}
	\label{fig:phase_separation1}
\end{figure}

The reduced adsorption of both components, defined as $\Gamma^{\prime}_i=\int_0^{\infty}( \rho_i(z)/\rho_i^\textmd{bulk}-1 )dz$, is depicted in Fig.~\ref{fig:1wall_pressure_gamma} (bottom panel) for systems S1 and S2. In both cases the behavior of big and small particles is very different for non-active systems, as $\Gamma^{\prime}_\textmd{b}>0$ (adsorption to the wall) and $\Gamma^{\prime}_\textmd{s}<0$ (depletion from the wall) for $a=0$. The difference between $\Gamma^{\prime}_\textmd{b}$ and $\Gamma^{\prime}_\textmd{s}$ shows a substantial reduction by increasing rate $a$. For $a>0.03$ the adsorptions finally converge to the same common value, which becomes practically constant with larger $a$, although the density profiles of both species still vary with $a$. 

All these results can be rationalized as follows:  If $a \ll 1$ the switching events are rare and the diffusive fluxes are still able to preserve the distinction between both components. However, for $a \gg 1$ many switching effects occur during the characteristic diffusive time $\tau_0=\sigma_\textmd{s}^2/D_\textmd{s}$, so particles do not have time to rearrange by diffusion, and they tend to experience the same average interparticle effective interaction ($u_{ij}(r) \rightarrow u_\textmd{eff}(r)$ for $a\rightarrow \infty$).

These arguments suggest that the inhomogeneous distribution of an equilibrium phase separated mixture should also be strongly affected by the activity, which we consider in the following. The equilibrium phase diagram of repulsive Gaussian mixtures has been extensively studied~\cite{Louis2000,Archer2001}. We select a mixture with $\epsilon_\textmd{bb}=\epsilon_\textmd{ss}=2$, $\epsilon_\textmd{bs}=1.888$, $\sigma_\textmd{b}=1.504 \sigma_\textmd{s}$ and $\sigma_\textmd{bs}=1.277 \sigma_\textmd{s}$. This choice of interaction parameters correspond to a system that undergoes fluid-fluid demixing above the critical point, located at $\rho_\textmd{T}^*\sigma_\textmd{s}^3=1.647$ and $x^*=0.7$. In order to ensure the phase separation of our Gaussian mixture, we chose a total number density given by $\rho_\textmd{T}\sigma_\textmd{s}^3=2.4$ and $x=0.5$, which is located well inside the unstable region. The mixture is confined inside a spherical cavity (cell) of radius $R=5\sigma_\textmd{s}$~\cite{Archer2005,Archer2005a}. This confinement is attained by means of repulsive spherically symmetric external potentials given by $\beta u_i^\textmd{ext}(r)=E_i (r/R )^{10}$ for $r \leq R$ and $\beta u_i^\textmd{ext}(r)=\infty$ for $r>R$, with $E_\textmd{b}=E_\textmd{s}=10$. 

The top-left plot of Fig.~\ref{fig:phase_separation1} shows the equilibrium  ($a=0$) density profiles for system S3, clearly exhibiting fluid-fluid phase separation in the cavity. Big particles are adsorbed or `condensed' close to the external wall of the cavity, whereas small ones are mainly distributed in the central region. This particular segregation is caused by the effective depletion attraction between the big particles and the wall induced by the smaller component~\cite{Archer2002}. The degree of separation between both components can be quantified by means of the inhomogeneity parameter $\delta$, defined as $\delta=\frac{4\pi}{xN_\textmd{b}+(1-x)N_\textmd{s}}\int_0^R \big| x\rho_\textmd{b}(r)-(1-x)\rho_\textmd{s}(r) \big| r^2dr$, where $N_\textmd{b}$ and $N_\textmd{s}$ are the (fixed) total number of big and small particles inside the cavity. $\delta=1$ implies that big and small particles completely demix into two non-overlapping regions, whereas $\delta=0$ means complete mixing. For $a=0$, we find $\delta=0.93$, indicating a high degree of demixing.

The steady-state profiles for larger activities are depicted in the rest of plots of Fig.~\ref{fig:phase_separation1} up to $a=10^3$ (details about the activity-induced out-of-equilibrium time evolution of the density profiles are shown in the Supplemental Material). The profiles show that $\rho_\textmd{b}(r)$ in the center of the cavity increases progressively with activity, while the height of the adsorption peak decreases. Conversely, increasing the conversion rate reduces $\rho_\textmd{s}(r)$ in the central region, narrowing the depletion layer of small spheres around the wall while propagating interesting intermediate peaks. In other words, increasing $a$ induces an activity-driven mixing and modified micro-structuring of both components. In particular, for activities above $a=1$, both density profiles approach each other, and finally for very large values of the activity, $a=10^3$, both profiles overlap, that is, demonstrate complete mixing. This is again the consequence of the fact that the system behaves as an effective one-size fluid in the limit of very fast exchange rates. The oscillations of the density profiles observed in this regime are the coordination layers of the effective one-component fluid induced by the repulsive interaction with the spherical wall.

The dynamic transition from a phase separated to a mixed fluid is clearly recognized in the inset of the bottom-right panel of Fig.~\ref{fig:phase_separation1}, where the inhomogeneity parameter $\delta$ is plotted as a function of the activity, tending to full mixing  $\delta\rightarrow 0$ for $a\rightarrow \infty$.  As observed, this transition is gradual and centered at $a=1$, which represents the inflection point separating mixed and demixed states. These results suggest that switching activity could be used as a tool to dynamically control the demixing state of mixtures, and are related with the previous observations in non-active systems that spinodal decomposition can be suppressed by chemical reactions~\cite{Glotzer1995}.

In summary, we demonstrated that active switching of pair interactions (induced, e.g., by conformational changes) has profound effects on the inhomogeneous microstructure of colloids. In particular, increasing activity suppresses the phase separation of in-equilibrium unstable systems inside cellular confinement, inducing a gradual mixing of the two components. This effect can be exploited for a non-equilibrium control of the degree of demixing and may have implications for liquid-liquid phase separation, condensation, and transient microstructuring in biological and synthetic functional cells~\cite{Shin,Berry2018} or non-equilibrium material design~\cite{Andreas,DNA_hydrogel,Heuser}.

\begin{acknowledgments}
The authors thank Upayan Baul for useful discussions. J.D. has received funding from the European Research Council (ERC) under the European Union's Horizon 2020 research and innovation programme (grant agreement No.\ 646659). A. M.-J. thanks the program “Visiting Scholars” funded by the University of Granada.
\end{acknowledgments}

\bibliography{DDFT-abbreviated}

\end{document}